# Non-user Inclusive Design for Maintaining Harmony of Real-Virtual Human Interaction in Augmented Reality

Chao Shi

Cygames Research, Cygames, Inc.
Tokyo, Japan
shi_chao@cygames.co.jp

## Abstract

Augmented reality enables the illusion of contents such as objects and humans in the virtual world co-existing with users in the real world. However, non-users who are not aware of the presence of the virtual world and dynamically move nearby might either cause a conflict by directly breaking into space where a user is talking to a Virtual Human (VH), or be troubled when try to avoid disturbing the user. To maintain harmony and keep both the user's and non-users' comfort, we propose a method that controls the VH to adjust its own position to avoid such potential conflict. The difficulty to address this problem is that the agent must avoid potential conflict in a natural way to keep the user away from feeling unnatural. Our idea is to endow the VH with three capabilities: anticipating non-users walking around, understanding how to establish and maintain proper formation to adapt to the environment, and planning to avoid conflicts by shifting formation in advance. We develop a non-user inclusive spatial formation model that realizes natural arrangement shift corresponding to the environment based on theoretical sources from literature. We implemented our proposed model into a VH behavior planning system to achieve natural conflict avoidance. Evaluation experiments showed that it successfully reduces potential conflicts caused by non-users.

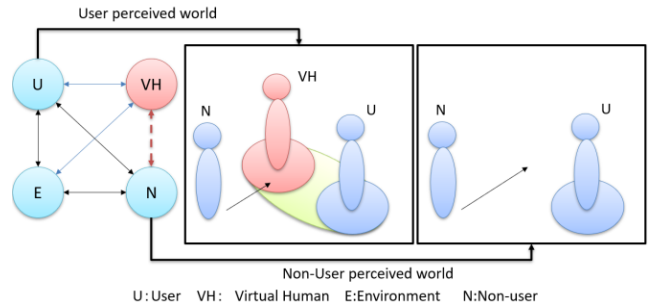

Figure 1: When interacting with a virtual agent in a public space AR environment, other non-user pedestrians might cause conflicts due to their perception limitations.

## Introduction

Augmented reality can integrate virtual contents into the real world while maintaining the user's sensations of the real world, shows great potential of use in our daily life. As the AR market size is expected to grow at a compound annual growth rate of 40%-85% in the next few years [1], it is proper to assume that AR displays will become a common sight not only in domestic environment such as homes but also in open public spaces like museums, shops and streets. In particular, in public spaces, AR technologies can have great promise of use in information providing using a virtual human (VH) as multimodal social user interface [2]. Many previous studies in AR focused on the seamless visual integration of virtual entities into the real world, such as by generating plausible lighting for photometric integrity and occlusion handling for geometric integrity. With such technologies, a VH can generate an illusion of physicality as it really co-exists with the user in the physical environment.

Recent researches on interacting with VHs in AR have shown that it is important for a VH to maintain spatial coherence with the environment and social coherence with real human, i.e., the user, through plausible behaviors [3]. However, when considering using such VHs in open public spaces, there are also non-users nearby. Unlike users who can observe VH through see-through devices, non-users are not aware of virtual entities that are superimposed upon the real world. Such asymmetric perception towards virtual content between user and non-users can cause problems that break the harmony. For instance, as shown in Fig.1, in a situation where a user is engaging in a face to face conversation with a VH, a non-user may pass directly between them or even through the VH's body, which will obviously break the common social/physical norms [4]. On the other hand, a non-user pedestrian who want to avoid disturbing an AR user may often be troubled by the inability to detect the interaction territory.

To the best of our knowledge, none of previous studies have investigated the problem of break of harmony caused by asymmetric perception between AR user and non-users and did not propose a method to address such situation.

In this study, we introduce a non-user inclusive design of a behavior planner of a user exclusively perceptible AR VH for maintaining harmony in public space where other non-users co-exist. Our approach is to use territories of the user and the user-VH group to predict future trajectories of non-user pedestrians and detect potential conflicts. Based on a model of arrangement adjustment that evaluates two important interaction comfort: (1) recognition and avoidance of potential conflict (outgroup comfort) (2) maintain the user's interaction comfort and presence (ingroup comfort), the VH can finally avoid conflicts while maintaining interaction with the user by natural spatial formation shifting behavior.

## Related Works

This section provides an overview of related work on VHs in MR and their interaction with physical entities in the real environment and the consideration of non-users.

### Virtual Humans in MR

A VH is generally defined as a computer graphic entity to simulate people's behavior that can be presented using display devices. They can either appear in a virtual environment or can share a physical space with real humans. VHs can be categorized as either avatars or agents based on the entity controlling them. Avatars are controlled by humans, while agents are controlled by computer programs [5].

The terms copresence and social presence are commonly used to discuss and reason about users' interactions with VHs. The term copresence was first defined by Goffman et al. [6] as existing when people sensed that they are able to perceive others and that others are able to actively perceive them. It has also been defined as a condition in which two-way human interactions can take place or a sense of "being there together". The term social presence was first defined by Short et al. [7] as the degree of salience of the other person in the interaction and the consequent salience of the interpersonal relationships. It has also been defined as "one person feels another person's presence" and "one's sense of being socially connected with the other". To overcome the shortcoming of such confusing and occasionally contradictory definitions, Skarbez et al. [8] proposed the use of the term Social Presence Illusion to explicitly refer to the illusory feeling of being together with and engaging with a real sentient being, and Copresence Illusion to refer to the feeling of "being together" in a virtual, mediated or real space.

Researchers have investigated factors that could influence users' sense of copresence illusion and social presence illusion. Previous studies have shown the importance of the traits of VHs, such as appearance realism [9], the naturalness of behaviors [10], the ability to exchange social cues (gaze, gestures, etc.). Many social science studies where users interacting with VHs in highly immersive social scenarios used VR environment to make the interaction experience as realistic as possible.

There is a relatively small amount of research attempting to bring VHs into users' real space using AR technology. In an immersive AR environment where VHs are superimposed upon the real world, previous knowledge about the interaction between human user and VHs in projection-based MR and VR could be partially generalized [9-13]. Since the VHs can move freely in the physical spaces, maintaining plausible physical-virtual cross world coherence could be more important in user perception of the VHs. For coherence between VH and environment, Kim et al. [3,4] conducted a series of studies indicated that a VH demonstrating awareness and behavioral adaption of the physical environment can elicit higher social presence ratings. Lee et al. [14] showed that a VH showing capability of physical influence has a positive impact on the user's sense of copresence, physicality and the VH's abilities.

### Consideration and Exploration of Non-users' Influence

There are some previous studies in the field of XR try to find and analyze the impact of non-users' presence in shared spaces. For example, in the field of VR, Yang et al. [16] proposed a system called ShareSpace that allows non-VR users to communicate their needs for physical space to the VR users wearing an HMD and immersed in their VR experience. Marwecki et al. [15] proposed a system called VirtualSpace that could overload multiple VR users engaging in their own virtual worlds into the same physical space [15]. While in AR environment, although users could maintain sensations of the real world, as their operation of the HMD device and interaction with virtual contents are not transparent to non-users, such asymmetric awareness might cause a cognitive disparity between them. Alallah et al. [17] contrasted performer's (user) and observer's (non-user) perspectives of social acceptability interactions with HMDs under different social contexts. They found both similarities and differences in social acceptability and suggested it is important to consider both perspectives when exploring social acceptability of emerging technologies.

In the field of intelligent agents, there are relatively more researches focused on interaction with physical agents in public spaces. For example, studies on how to navigate a social robot considering social norms to respect and not disturb others in the environment [18], how to deal with congestion caused by the robot [19] or problem of children's abuse [20], etc. These studies are based on the situation that physical agents can be symmetrically perceived by both user and non-users in the environment, thus the problems and solutions are very different from which in AR situation. Despite the differences in the specific situations, the final goal that maintains harmony among users, agents and other non-users in the environment, is essentially common.

# A Virtual Human that Maintain Harmony in Public Augmented Reality

To maintain the harmony the VH should consider both the user and non-user's comfort. Our general idea is to make the agent adjust its position to avoid potential conflict. As the user is mainly paying attention to the agent in the interaction, the agent's position adjustment behavior could lead the user's gaze direction away from the non-user pedestrian's future trajectory and thus maintain their comfort simultaneously. The difficulty is the agent needs to perform it in a natural way to avoid giving users unnatural feelings that might reduce its social presence. Previous literature [21] reported that dyadic groups adjust their spatial arrangement to adapt to changes in the external environment. Thus, we believe that by intimating such arrangement adjustment behavior a VH could realize a natural way to avoid potential conflicts. Our idea is as below: first, predict future trajectories of non-users based on what they could perceive, i.e., the user's individual territory, then combining the territory of the user-VH dyadic group to detect potential conflicts; when potential conflict detected, the VH achieve natural conflict avoidance by intimating the arrangement adjustment behavior in response to changes in the external environment.

## Modeling of the Conflict Avoidance Behavior

**Territory of the user:** The proper distances which human keep with others is driven by psychological factors and social norms related to the concept of personal space (PS): the area surrounding individuals, into which strangers' intrusion would cause discomfort [22]. We use PS to describe the territory of the user and simply used its commonly represented shape as a circular area surrounding the user.

**Territory of user-VH Group:** We used the sociological concept of F-formation from the works of Ciolek and Kendon. F-formation describes geometric patterns that people tend to form during social interactions. Moreover, in [21] the authors defined six spatial arrangements and further defined them into three patterns as vis-à-vis (closed), L-shaped (semi-open), and side-by-side (open) based on the openness of arrangement which depends on the interlocu-tors' pose (position and orientation). These dyadic arrangements are not used equally, but often selected depend very much on what people are doing, and on who they are in terms of their personal and interpersonal characteristics. Ciolek [21] suggested the use of arrangement also influenced by spatial context of the encounter and further reported that the degree of openness of spatial arrangement in a dyadic conversation is influenced by both the definiteness of a given place and the crowdedness (the intensiveness of pedestrian flow). In this paper, we use F-formation model to describe the territory of the user-VH dyadic group.

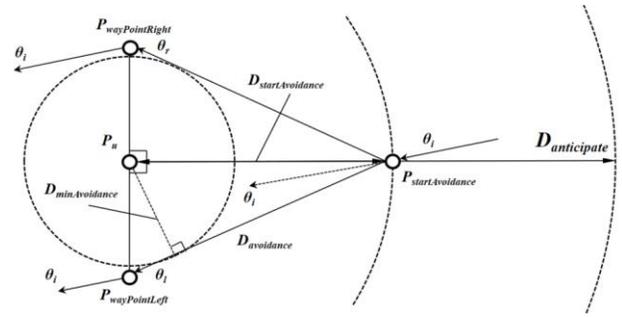

Figure 2: Prediction of pedestrian's trajectory.

**Prediction of non-user pedestrian's trajectory:** For pedestrians who are dynamically walking, a notion of pedestrian personal space (PPS) was used and reported to have a much larger range on the frontal side [23]. Pedestrians not only avoid intruding others' PS but also try to keep others away from their PPS by adjusting the walking trajectory. As a result, when pedestrians find there is a stopped person right on their walking direction, they start to adjust walking trajectory from a start avoidance distance and then pass by within a minimum avoidance distance [24]. We used these two distances to predict the future trajectory of a pedestrian who is walking towards the user from a long distance.

In public space, the VH only need to anticipate pedestrians walking nearby. We used the concept of c-space of F-formation, which is used for interlocutors in a formation to observe pedestrians nearby to define the anticipate distance $D_{anticipate}$ and set it as 6m according to [21]. We simply used the pedestrians walking trajectories to predict their future positions as below:

$$EP_i(t) = P_{current} + v_i \cdot t \tag{1}$$

We then further calculate the interpersonal distance between the user and the pedestrian on each future position. A pedestrian with a minimum interpersonal distance $d_{min}$ with the user larger than the minimum avoidance distance $D_{minAvoidance}$ does not need to change the walking trajectory. On the opposite, as shown in Fig.2, for the pedestrians with $d_{min}$ smaller than $D_{minAvoidance}$, they need to adjust walking trajectories to avoid the user. According to [24], a pedestri-an would start to change his/her direction from a start avoidance distance $D_{startAvoidance}$ and passed the user while keeping the $d_{min}$ no less than $D_{minAvoidance}$. We modeled it as follow: when the pedestrian approaches the user within $D_{startAvoidance}$, he/she changes the course and walks toward a waypoint $P_{wayPointLeft}/P_{wayPointRight}$ on either left or right side that keeps his/her $d_{min} = D_{minAvoidance}$. After reaching the waypoint, the pedestrian then returns to original direction and walks away. During the avoidance, the future position of the pedestrian could be calculated by using the walking direction towards waypoint $\theta_{avoidance}$ and the $D_{avoidance}$ which represents the distance from $P_{startAvoidance}$ to the waypoint. $D_{avoidance}$ is calculated as below:

$$D_{avoidance} = D_{startAvoidance}/\cos Angle(\theta_{pi,pu}, \theta_{avoidance}) \quad (2)$$

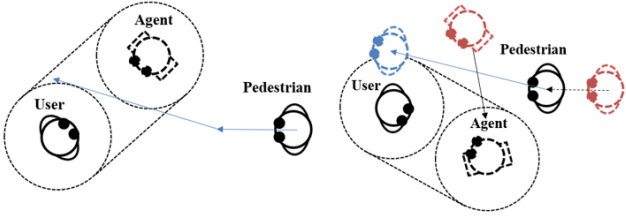

Figure 3: Potential conflict prediction and conflict avoidance

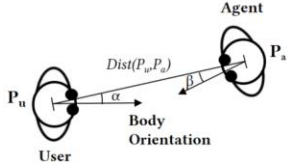

Figure 4: Calculation of Spatial Arrangements used in F-formation

Where $Angle(\theta_{pi,pu}, \theta_{avoidance})$ is the angle between the direction from $P_{startAvoidance}$ to the position of the user $P_u$ and the direction from $P_{startAvoidance}$ to a waypoint and could be calculated as below:

$$Angle(\theta_{pi,pu}, \theta_{avoidance}) = \arcsin D_{minAvoidance}/D_{startAvoidance} \quad (3)$$

And then the walking direction toward each waypoint $\theta_l/\theta_r$ could be obtained.

**Comfort Model:** As shown in Fig.3, when the predicted future position of a pedestrian intrudes the territory of user-VH group, it means there is a potential conflict. In such case, the VH needs to adjust its position to where it can both maintain proper F-formation with the user and keep the pedestrian away from intruding the formation. We define the former as the user-VH's In-Group Spatial comfort, and the latter as Out-Group Spatial Comfort.

For Out-Group Spatial Comfort, we used a distance-based comfort model proposed and successfully used in [19] to evaluate whether the VH could avoid the conflict during moving to a position $P_x$. The idea of this model is that it is more comfortable for the user if the distances to nearby persons are larger. Based on this model, the user's comfort based on the influence of pedestrians nearby at time $t$ is defined as:

$$c_{dist}(c \rightarrow P_x, S, t) = \min_{i \in S}(a/d(g,i,t)+b) \quad (4)$$

Table 1: Preference for each spatial arrangement corresponded to spatial context

|  | Closed | L-shaped | Open |
|---|---|---|---|
| Open-Space / Uncrowded | High | Middle | Low |
| Open-Space / Crowded | Low | High | Low |
| Near-Wall / Uncrowded | Middle | Middle | High |
| Near-Wall / Crowded | Low | High | Middle |

Where $d(g,i,t)$ is the distance between pedestrian $i$ and the dyad $g$. $g$ is a line segment between the position of the user and the VH. $S$ represents set pedestrians walking near $g$, and $a$ and $b$ are regression parameters.

The overall Out-Group Comfort is equal to its smallest value during the pedestrians passing, thus it could be calculated as below:

$$comfort_{OutGroup}(c \rightarrow P_x, T) = \min_{t \in T} c_{dist}(c \rightarrow P_x, S, t) \quad (5)$$

$T$ is the segments in which all the pedestrians nearby will leave the dyad's c-space. We bounded the value to be in the range [0, 1], and used the distance of imitate space and $D_{minAvoidance}$ for regression analysis and yielded the parameters a=-1370.25, b=3.045.

For In-Group Comfort, as shown in table 1 [21], people have a different preference towards each arrangement under different environment context. As shown in Fig.4, α/β is the angle between the body orientation of the user/agent and the vector from the user/agent to the interlocutor. The model first checks at a position $P_x$ whether a VH could maintain F-formation with the user as below:

$IsFformationAvailable =$

$$\begin{cases} 1 & 0.6\,\text{m} \leq Dist(P_u, P_a) \leq 1.5\,\text{m} \text{ and } \alpha \leq 90\,\text{deg} \\ 0 & (otherwise) \end{cases} \quad (6)$$

and list up all the possible arrangement based on the equation as follow:

$$Type\,of\,Arrangement = \begin{cases} Closed & 0\,\text{deg} \leq (\alpha+\beta) \leq 60\,\text{deg} \\ L-shaped & 60\,\text{deg} < (\alpha+\beta) < 120\,\text{deg} \\ Open & 120\,\text{deg} \leq (\alpha+\beta) \leq 180\,\text{deg} \end{cases} \quad (7)$$

We define the In-Group Comfort model as below:

$$comfort_{InGroup}(c \rightarrow P_x, T) = c_{SpatialContext}(c \rightarrow P_x, T) \quad (8)$$

where $c_{SpatialContext}$ is the comfort based on the environment. To get the value of $c_{SpatialContext}$ for a candidate position $P_x$, recognition of definiteness and crowdedness of the environment and evaluation of formation generation is needed. We used the size of PS as the threshold value to define the territory of the dyad for spatial context, i.e., definiteness and crowdedness, and then calculate the distance between the pedestrians/walls (or another big object) to the dyad.

In addition, to establish an F-formation, the VA needs to set its orientation properly to keep $\beta \leq 90$ deg.

When an F-formation is available, the type of arrangement is evaluated by equation 1. Since the VA could control its own body orientation, based on the threshold of $\beta$, an arrangement that could be established at $P_x$ could be obtained.

The $c_{SpatialContext}$ could be calculated as below:

$$c_{SpatialContext}(c \rightarrow P_x, T) = \max(p \cdot c_{formation}) \tag{9}$$

where $p$ is the value that represents the influence of environment and be set accroding to Table 1 (high = 1.0, middle = 0.6, low = 0.2).

The overall utility for a target position $P_x$ is calculated below:

$$Utility(c \rightarrow P_x, T) = \tag{10}$$
$$(comfort_{InGroup}(c \rightarrow P_x, T) + c \cdot comfort_{OutGroup}(c \rightarrow P_x, T)) \Big/ 1 + movedist \cdot d$$

Where $c$ and $d$ are coefficients. Finally, among all the candidates of target positions, the VA chooses to move to the candidate position with the highest utility as the arrangement adjustment plan to be executed:

$$U_{decision} = \max(Utility(c \rightarrow P_x, T)) \tag{11}$$

## Public Space Conflict Avoidance Simulation

We conducted an analytic study based on simulation to evaluate the performance and generalizability of our proposed conflict avoidance model. We firstly conducted an ablation experiment to calibrate the parameters in a fixed environment setting, and then changed the environmental factors to investigate the influence of each factor.

### Ablation Experiment

#### Material and Methods

We created a system to simulate the situation where an AR user and a VH coexists with pedestrian non-user in open public space with Unity engine. To make the simulation environment as real as possible, we refer to the Vittorio Emanuele II Gallery Dataset [25] which is an annotated video about pedestrian and group dynamic proxemic behavior. The data was collected on the 24th of November 2012 (on Saturday afternoon, from 2:50 pm to 4:08 pm). The Walkway Level of Service Criteria (LOS) [26] of the dataset is LOS B, which represents low density and flows with minor conflicts (0.08-0.27 people/squared meter, 7-23 pedestrian/minute/meter). As shown in Fig.5, we created an open square environment with the same size (12 m x 12

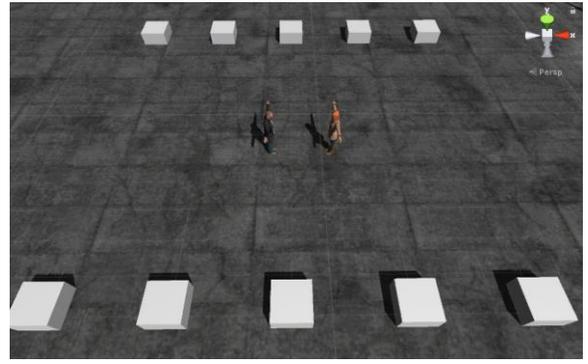

Figure 5: Environment Setting for Ablation Experiment.

m) and set the density of walking pedestrian as 0.25 people/square meter.

We put two virtual characters in the middle of the environment and set one as the user and the other as a VH. To create pedestrian flow, the system generates multiple characters as pedestrians and let them walk toward a goal in the environment. We put five boxes on each of the up and down sides of the environment and let each pedestrian randomly choose one as his/her destination at the beginning. When a pedestrian reaches a goal, he/she would then randomly choose a new goal from the five boxes on the other side, and thus we could create a random flow of pedestrians. In the simulation, all the pedestrians would avoid intruding the user's personal space following the trajectory described in Fig. 2. To simulate the potential conflicts in AR situation, we set the VH character invisible to the pedestrian characters and disabled its collision asset so that pedestrians would not avoid it and could pass through it.

The walking speed of the pedestrians is randomly set ranged from 1.0 m/s to 1.5m/s. The max moving speed of the VH is set to 1.5 m/s.

#### Factors

**Coefficient $c$.** As shown in equation 10, the coefficient $c$ is an application-dependent parameter, one should calibrate it to yield a reasonably good performance balancing both ingroup and outgroup comfort.

**Coefficient $d$.** Coefficient $d$ is used to weight the cost of VH's move distance in its avoidance behavior.

**User-VH Interpersonal Distance.** The proper interpersonal distance between interlocutors in a conversation depending on several factors such as culture, age, gender, personality and social context of interactions. One should set this parameter properly based on their own situation and scenario.

**Tracking Distance.** In real-world AR, a VH need to collect environmental information from sensors for its prediction and avoidance to potential conflicts. In the simulation, we use this parameter to simulate the sensing capability of different sensors to investigate its influence on the performance of the proposed model.

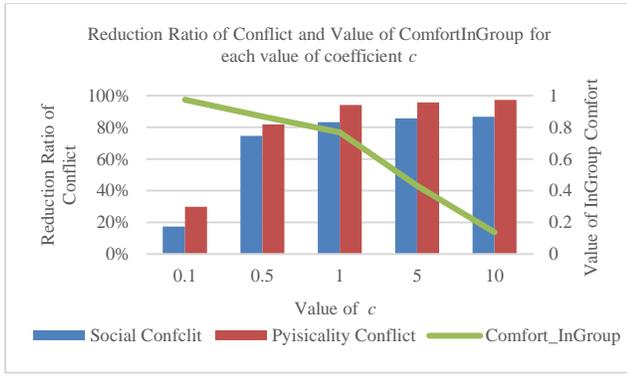

Figure 6: Influence of coefficient *c*

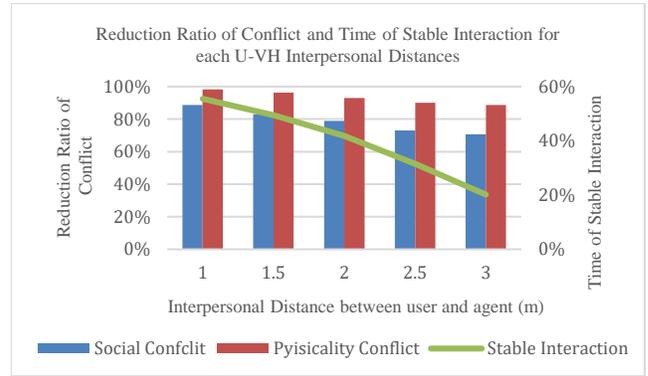

Figure 8: Influence of U-VH interpersonal distance

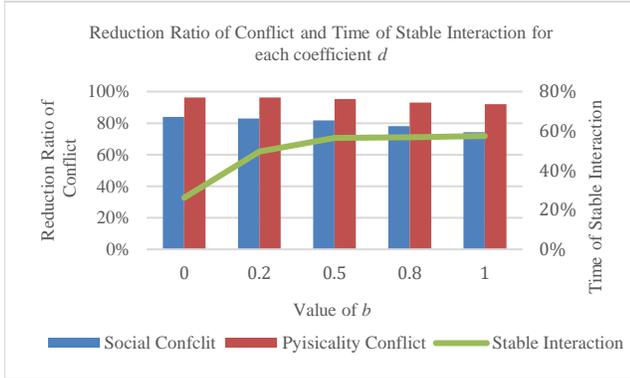

Figure 7: Influence of coefficient *d*

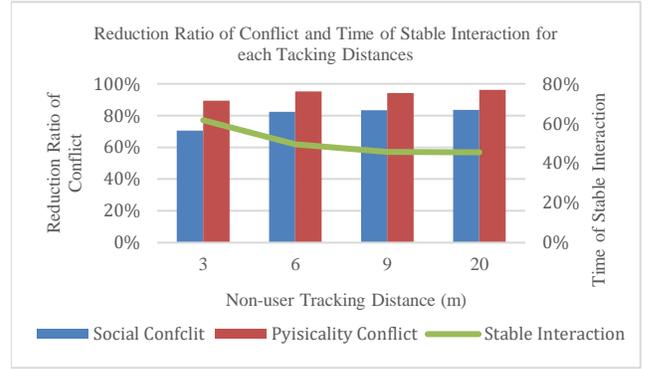

Figure 9: Influence of tracking distance

**Apparatus and Procedure**

We preliminarily set the initial value for each factor, and then calibrate them one by one. Initial value of *c* is set to 1, and *d* is set to 0.5. Based on social literature [22], initial value of U-VH interpersonal distance is set to 1.5 m, and tacking distance is set to 6 m. We selected one of the factors to calibrate while keeping the others unchanged with the initial values.

**Dependent Variables**

**Value of Ingroup Comfort.** We analyzed this variable for only coefficient *c* as its value could have a strong influence on the user-VH ingroup comfort.

**Detected conflicts and reduction ratio of conflict.** Since we are interested into what extent our proposed method could reduce conflicts caused by non-user which could be used to evaluate the effectiveness and generalizability of the model, we recorded the count of detected conflicts in each simulation trial for further analysis. We calculate the reduction ratio of conflict avoidance as below:

$$ReductionRatio = \frac{DC_{noneAvoidance} - DC_{Avoidance}}{DC_{noneAvoidance}} \quad (12)$$

In addition, we further define the conflicts into two types: (1) social conflict, which represents that non-users intrude the territory of the user-VH group; (2) physicality conflict, which represents that non-users passing through the body of the VH by occupying same space with it.

**Time percentage of stable interaction.** Although the VH's conflict avoidance behavior aims to maintain harmony among the user, non-user and the VH, too frequent formation adjustment behaviors might be perceived as unnatural and thus might have a negative impact on the user's interaction experience with the VH. Thus, we used the simulation system to record time length of both agent's formation adjusting period and stable interaction period. Then we could further analyze how the changes of environment influence the agent's behaviors.

**Results**

Figure 6-9 shows the results. We firstly calibrate coefficient *a*. The result showed that as *c* grows, the reduced ratio of conflict avoidance increases while the *comfort-InGroup* decreases. We set *c* as 1 for its good balance for both the two dependent variables. For coefficient *d*, the result showed that it decreases the reduction ratio of conflict

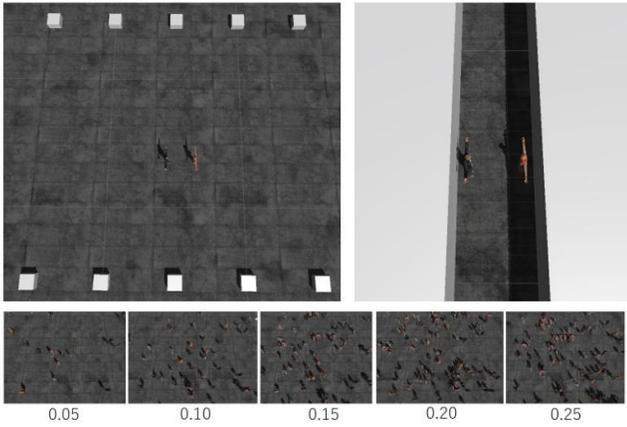

Figure 10: Environment Setting for the factors of breadth and density.

avoidance and increases the time of stable interaction when it grows. We finally set $d$ as 0.5. For U-VH interpersonal distance, the results showed that as the distance grows larger, it would lower both the two dependent variables. When using AR HMD interacting with a VH, the possible interpersonal distance is largely based on the field of view of the device. This result suggests that using an HMD with larger FOV, the proposed model could achieve better performance. While for the last variable the tracking distance, we can see that as the distance increases to about 6 m, the model could achieve a good performance. A much larger tracking distance such as 20 m would not significantly impact the performance. Thus, we consider with our proposed model, a tracking distance of about 6 m is sufficient for use in a real-world situation.

## Performance Evaluation in different environments

### Material and Methods

Using the parameters calibrated in the ablation experiment, we evaluated the proposed model in different environments. the situation where AR user and VH coexist with pedestrian non-user in open public space with Unity engine.

### Factors

**Breadth.** We created two environments with different breadth for the test (Figure 10):
A. A 20m x 20m open spacious square environment.
B. A 3m x 20m narrow passage environment. It is representative of a common corridor.

**Density.** For each Environment, we controlled the simulator's number of pedestrians to create flows with different densities, ranging from free flow to minor conflicts scenes. We selected 5 Density levels (0.05 to 0.25 pedestrians per square meter, by 0.05 p/m$^2$ step).

**Conflict Avoidance.** Our goal is to test to what extent the proposed model could reduce potential conflict caused by non-user pedestrians, thus we used 2 conditions for conflict avoidance behavior: none and our proposed method.

### Apparatus and Procedure

We ran the simulation for 20 trials: 2 Environment x 5 density x 2 conditions. Each simulation trial lasted 10 minutes. We were interested in both the effectiveness of the conflict avoidance behavior and the stability of the user-VH conversation and therefore analyzed the count of detected conflicts and time ratio of user-VH stable interaction.

### Dependent Variables

**Detected Conflicts and Reduction ratio of conflict.** Since we are interested into what extent our proposed method could reduce conflicts caused by non-user which could be used to evaluate the effectiveness and generalizability of the model, we recorded the count of detected conflicts in each simulation trial for further analysis.

**Time Percentage of Stable Interaction.** As the same as in the ablation experiment.

### Results

Table 2 and Table 3 shows the results in the two environments with different breadth. Numbers in parentheses represent the number of detected conflicts in the two conditions of conflict avoidance behavior. Results showed that in both of the two environments, with the proposed model a VH could completely avoid all the potential conflicts while maintaining a relatively stable interaction for low density (0.05). As the pedestrian density increases, the two kinds of conflicts would also increase correspondingly, while the number of social conflicts increases more rapidly than that of physicality conflicts. We can also see that the VH need to the user more time for the arrangement adjustment behavior to avoid conflicts and maintain harmony. The time percentage of stable interaction significantly reduces as the density grows larger. For comparison between the two environments, from the result, we found that the increase of detected conflicts and decrease of time percentage of stable interaction grows slower in the narrow passage environment than those in the open spacious square environment. The main reason is in such narrow space, the walking trajectory of pedestrians are relatively stable, and thus the VH could "hide" behind the user and user the user's personal space as a cover to reduce potential conflict.

Table 2: Reduce ratio of conflicts and time percentage of stable interaction for each density in open spacious square environment

| Density | 0.05 | 0.10 | 0.15 | 0.20 | 0.25 |
|---|---|---|---|---|---|
| Reduce ratio of social conflict | 100% (0/86) | 91% (18/204) | 82% (54/306) | 81% (78/410) | 80% (104/522) |
| Reduce ratio of physicality conflict | 100% (0/62) | 100% (0/150) | 95% (9/188) | 91% (24/272) | 91% (31/350) |
| Time percentage of Stable interaction | 77% | 64% | 50% | 31% | 21% |

Table 3: Reduce ratio of conflicts and time percentage of stable interaction for each density in narrow passage environment

| Density | 0.05 | 0.10 | 0.15 | 0.20 | 0.25 |
|---|---|---|---|---|---|
| Reduce ratio of social conflict | 100% (0/67) | 97% (3/105) | 94% (10/177) | 83% (38/222) | 75% (74/297) |
| Reduce ratio of physicality conflict | 100% (0/24) | 100% (0/78) | 99% (1/95) | 97% (4/131) | 97% (5/189) |
| Time percentage of Stable interaction | 97% | 94% | 82% | 70% | 63% |

The above results suggested that our proposed model could be effectively used in both wide and narrow environments in low density (under 0.15) of pedestrians. While in more crowed public spaces, some other methods need to be considered.

## Conclusion

In public space where an AR user interacts with a virtual human, other non-users could often cause conflicts that break the harmony due to the asymmetric perception between them. We addressed this problem using a spatial formation arrangement adjustment behavior for a virtual human in AR to automatically avoid such potential conflict. We proposed a computational model and evaluated its performance in a simulation test. The study demonstrates the usefulness and generalizability of the proposed model.